\documentstyle[prl,aps,twocolumn]{revtex}

\begin{document}

\title{
{\normalsize\begin{flushright}
UB-ECM-PF-98/08,
ITP-UH-04/98,
UTTG-04-98\\
hep-th/9803196,
to appear in Phys. Rev. Lett.\\[1.5ex]
\end{flushright}}
D-String on Near Horizon Geometries and
Infinite Conformal Symmetry
}
\author{
Friedemann Brandt\,$^a$, Joaquim Gomis\,$^{b,c}$, Joan Sim\'on\,$^b$}

\address{
$^a$ Institut f\"ur Theoretische Physik, Universit\"at Hannover,
Appelstra\ss e 2, D--30167 Hannover, Germany\\
$^b$ Departament ECM,
Facultat de F\'{\i}sica,
Universitat de Barcelona and 
Institut de F\'{\i}sica d'Altes Energies,
Diagonal 647,
E-08028 Barcelona, Spain\\
$^c$ Theory Group, Department of Physics,
University of Texas, Austin, TX 78712, USA
\\[1.5ex]
\begin{minipage}{14cm}\rm\quad
We show that the symmetries of effective D-string actions in constant
dilaton backgrounds are directly related to homothetic motions of the
background metric. In presence of such motions, there are infinitely
many nonlinearly realized rigid symmetries forming a loop (or loop like)
algebra. Near horizon (AdS) D3 and D1+D5 backgrounds are discussed in 
detail and shown to provide 2d interacting field theories with infinite 
conformal symmetry.
\\[1ex]
\end{minipage}
}

\maketitle

The recent past has seen an increasing interest in 
the conjecture of a correspondence between large 
$N$ limits of certain $d$-dimensional conformal field theories 
and supergravity on the product of
$(d+1)$-dimensional anti-de Sitter ($AdS$) space with a compact manifold
\cite{m,witten}. This suggested to
consider world-volume brane actions on near horizon backgrounds.
M2, M5 and D3 branes have been studied \cite{m,k0,k1}
and interacting 
$(p+1)$-dimensional theories in Minkowski space-time with conformal
$SO(2,p+1)\times SO(d-p-1)$ symmetry were found \cite{k2}.
The conformal symmetries of these branes reflect
the isometries of $AdS_{p+2}\times S^{d-p-2}$.
The case of a D-string in the near horizon geometry of a
D1+D5 brane was also considered in \cite{k2}.

In this work we study the rigid symmetries of effective D-string
actions of the Born-Infeld type on curved
backgrounds with constant dilaton.
We find that the symmetries are related with
homothetic motions of the background metric. Each of these motions
gives rise to infinitely many nonlinearly realized rigid symmetries,
with the Born-Infeld gauge field transforming in a non-trivial way.
The algebra of these symmetries is a loop generalization of the algebra
associated with the homothetic motions.
We spell out the symmetry transformations before gauge fixing and in the
static gauge for the world-sheet diffeomorphisms. The gauged fixed 
transformations generate infinitely many symmetries of interacting
(1+1)-dimensional field theories in a flat space-time.

We then specify these general results for particularly
interesting D3 and D1+D5 brane backgrounds and show that the
gauge fixed field theories in the respective near horizon
($AdS$) backgrounds have infinite conformal symmetry.
In the case of the D3 background the symmetry group is a loop
generalization of $ISO(1,3)\times SO(6)$.
In the near horizon limit there is
an enhancement of the symmetry to the loop generalization of
conformal $SO(2,4)\times SO(6)$ due to the $AdS$ geometry. 
The symmetry group contains as a subgroup a loop version 
of conformal $SO(2,2)$ with nonlinearly realized special
conformal transformations. 

In the case of a D-string on a near
horizon D1+D5 background we get an interacting theory with
infinite conformal
$SO(2,2)\times SO(4)\times ISO(4)$ loop symmetry.
The zero modes of the loop algebra reproduce the 
corresponding results of \cite{k2}.

We remark that these structures are not restricted to
Dirac-Born-Infeld actions. Rather, they are present in a
more general set of models studied here.
Hence, in appropriate backgrounds one gets a set
of conformal field theories.
This does not exclude that kappa-invariant extensions
of our formulation and/or T-duality properties may select the 
Dirac-Born-Infeld action.

It is natural to wonder how these results extend to D$p$-branes
with $p>1$. This is not known; a complete classification of the
symmetries for $p>1$ has not been carried out so far.
Of course, the presence of infinitely many symmetries may
well be restricted to the case $p=1$, as the two-dimensional
case is often special. On the other hand, the presence
of a Kac-Moody version of the conformal group $SO(2,4)$
for D3 branes in the near horizon 
geometry has been conjectured recently in \cite{p} and
would be reminiscent of our result for $p=1$. Work in this
direction is in progress.

{\em Symmetries and homothetic motions.}
The effective Born-Infeld actions for D-strings considered here
can be cast in a form
similar to the familiar sigma model formulation
of the Nambu-Goto action. In this form
they are contained in a more general class of models with an 
action of the form
\begin{eqnarray}
 S =  \frac{1}{2} \int d^2\sigma\left\{
\sqrt{\gamma}\, \gamma^{\mu\nu}
f(\varphi)\, g_{mn}(x)
\partial_{\mu}
x^m\partial_{\nu}x^n 
\right. \nonumber \\
 \left. + \epsilon^{\mu\nu}[b_{mn}(x)\partial_{\mu}
x^m\partial_{\nu}x^n + D(\varphi)\, F_{\mu\nu}]
\right\} 
\label{density}
\end{eqnarray}
where $\gamma_{\mu\nu}$ is an auxiliary world-sheet metric,
$\varphi$ is an auxiliary scalar field,
$\epsilon^{\mu\nu}$ the usual Levi-Civita tensor density, and
$F_{\mu\nu}=\partial_\mu A_\nu-\partial_\nu A_\mu$ 
is an abelian field strength. 
$g_{mn}$ and $b_{mn}$ are to be
thought of as target space metric and 2-form
respectively. We do not impose restrictions on $f$ and $D$ 
apart from $f,D\neq constant$, but we note that 
one of them may be chosen 
conveniently (only the relative choice of $f$ and $D$ 
characterizes a particular model).
Born-Infeld actions arise for
\begin{equation} 
f^2(\varphi)-D^2(\varphi)=1.
\label{bi}\end{equation} 
Indeed, eliminating the auxiliary fields
$\gamma_{\mu\nu}$ and $\varphi$ using the equations of motion,
the Lagrangian turns for (\ref{bi}) into
\begin{eqnarray}
& L_{BI}=
\sqrt {-\det({\cal G}_{\mu\nu}+F_{\mu\nu})}+
\frac 12 \epsilon^{\mu\nu}{\cal B}_{\mu\nu}
\nonumber\\
& {\cal G}_{\mu\nu}=g_{mn}(x)\partial_\mu x^m \partial_\nu x^n,\
{\cal B}_{\mu\nu}=b_{mn}(x)\partial_\mu x^m \partial_\nu x^n. &
\label{BI}
\end{eqnarray}
This represents Born-Infeld models with a ``WZ-term''
determined by $b_{mn}$
and a constant dilaton which may be made explicit by
rescaling $g_{mn}$ and $A_\mu$. More general 
Born-Infeld models, in particular models with 
nonconstant dilaton, can also
be cast in a sigma model form \cite{ours1,ours2}, but
are not considered here.

In \cite{ours2} we have shown among others that all the
rigid symmetries of actions (\ref{density}) (and generalizations
thereof) are determined by generalized Killing vector equations.
An analysis of these equations, similar to the one performed for
the example treated in \cite{ours1}, shows
that the rigid symmetries of
models (\ref{density}) are generated by
\begin{eqnarray}
\Delta x^m &=& \xi_i^m(x)\, \lambda^i(\varphi),
\quad \Delta \gamma_{\mu\nu}=0,
\nonumber\\
\Delta \varphi &=& K_i\, \lambda^i(\varphi)\, f(\varphi)/f'(\varphi),
\nonumber\\
\Delta A_\mu &=& (1/D')
[ f\,\sqrt \gamma\, \epsilon_{\mu\nu}
\gamma^{\nu\varrho} g_{mn} (\Delta x^n)' \partial_\varrho x^m
\nonumber\\
& &+(b_{nm}\Delta x^n +\lambda^i q_{mi})'\partial_\mu x^m
 -A_\mu (D'\Delta\varphi)'].
\label{SOL}\end{eqnarray}
Here prime denotes differentiation with respect to
$\varphi$, the $\lambda^i(\varphi)$ are {\em arbitrary} functions
of $\varphi$, and $\{\xi^m_i(x),q_{mi}(x)\}$ denotes a 
complete set of inequivalent solutions of
\begin{eqnarray}
& {\cal L}_i\, g_{mn}(x) = -K_i\, g_{mn}(x),\quad K_i=constant &
\label{homo1}\\
& {\cal L}_i\, b_{mn}(x) = \partial_n q_{mi}(x)-\partial_m q_{ni}(x) &
\label{homo2}
\end{eqnarray}
where ${\cal L}_i$ is the Lie derivative along $\xi_i$.
Using (\ref{homo1}) and (\ref{homo2}),
it is not difficult to verify that the above transformations $\Delta$
generate indeed symmetries of an action (\ref{density}).

The symmetries of Born-Infeld actions
(\ref{BI}) are obtained from the above
formulas by eliminating the auxiliary fields $\gamma_{\mu\nu}$ and
$\varphi$, resulting in
\begin{eqnarray}
\Delta x^m&=& \xi_i^m(x)\, \lambda^i({\cal F})
\nonumber\\
\Delta A_\mu&=& 
\left[V_{\mu i}(1-{\cal F}^2)+W_{\mu i}(1-{\cal F}^2)^{3/2}\right]
\frac{d\lambda^i({\cal F})}{d{\cal F}}
\nonumber\\
& &+A_\mu K_i\left[{\cal F}^{-2}-2+({\cal F}-{\cal F}^{-1})
\frac{d}{d{\cal F}}\right]\lambda^i({\cal F})
\label{z1}
\end{eqnarray}
where
\begin{eqnarray}
V_{\mu i}&=&-\sqrt{{\cal G}}\,\epsilon_{\mu\nu}{\cal G}^{\nu\varrho}
\xi_i^m(x)g_{mn}(x)\partial_\varrho x^n
\nonumber\\
W_{\mu i}&=&[b_{mn}(x)\xi_i^n(x)-q_{m i}(x)]\partial_\mu x^m
\nonumber\\
{\cal F}&=&
\frac 12\,{\cal G}^{-1/2}\epsilon^{\mu\nu} F_{\mu\nu}\, ,\
{\cal G}=-\det ({\cal G}_{\mu\nu}).
\label{z2}
\end{eqnarray}

Let us now comment on the nature of the above symmetries.
The occurrence of arbitrary functions $\lambda^i(\varphi)$
in (\ref{SOL}) implies that each nontrivial solution to 
(\ref{homo1}) and (\ref{homo2}) gives rise
to {\em infinitely many rigid symmetries}.
Equation (\ref{homo1}) defines so-called homothetic motions of $g_{mn}$ 
and the $K_i$ are called homothetic constants \cite{yano}.
Homothetic motions with nonvanishing homothetic constants
are called proper because the others
are just isometries of the metric. One can always
choose a basis of homothetic motions such that at most one of them
is proper. Without loss of generality, we can thus use $i=1,2,\dots$ 
for isometries of the metric, reserve $i=0$
for a proper homothetic motion (if any), and
normalize $\xi_0$ such that $K_i= \delta_i^0$.

The commutator of a proper homothetic motion and
an isometry of the metric is always again an isometry, as
(\ref{homo1}) implies $[{\cal L}_0,{\cal L}_i]g_{mn}=0$. 
The algebra of homothetic motions 
is thus of the form
\begin{equation}
[{\cal L}_i,{\cal L}_j]={c_{ij}}^k {\cal L}_k\ ,\quad
[{\cal L}_0,{\cal L}_i]={c_i}^j {\cal L}_j
\quad (i,j,k\geq 1)
\label{homo3}\end{equation}
where ${c_{ij}}^k$ and ${c_i}^j$ are structure constants.

The presence of arbitrary functions of $\varphi$ in (\ref{SOL})
(which turn into functions of ${\cal F}$ upon elimination of $\varphi$)
implies that the algebra of the corresponding symmetries is a loop version
of (\ref{homo3}), the role of the loop variable being played by
$\varphi$ (or a function thereof).
This is seen by expanding the functions $\lambda^i$ in
a suitable basis for functions of $\varphi$. A 
particularly nice form
of the algebra emerges in a basis
consisting of powers of the function $f(\varphi)$ 
occurring in (\ref{density}).
We denote the corresponding basis of symmetries
by $\{\Delta^\alpha_i\}$ where $\alpha$ indicates the power of
$f(\varphi)$,
\begin{eqnarray}
\Delta^\alpha_i x^m = -\xi^m_i(x) f^\alpha(\varphi),\,
\Delta^\alpha_i \varphi = -\delta^0_i\,
\frac{f^{\alpha+1}(\varphi)}{f'(\varphi)}\, .
\label{homo4}
\end{eqnarray}
It is now straightforward to verify
that in this basis the symmetry algebra reads on $x^m$ and $\varphi$
\begin{eqnarray}
{} [\Delta^\alpha_i,\Delta^\beta_j] &=& 
{c_{ij}}^k \Delta^{\alpha+\beta}_k \qquad (i,j,k\geq 1)
\label{homo5}\\
{} [\Delta^\alpha_0,\Delta^\beta_i] &=& 
({c_i}^j-\beta\delta_i^j) \Delta^{\alpha+\beta}_j 
\qquad (i,j\geq 1)
\label{homo6}\\
{} [\Delta^\alpha_0,\Delta^\beta_0] &=& 
(\alpha-\beta)\Delta^{\alpha+\beta}_0\ .
\label{homo7}
\end{eqnarray}
Note that (\ref{homo5}) is a loop algebra associated with
the isometries of the metric. Hence, if there is no
proper homothetic motion, the symmetry algebra is
a true loop algebra. In presence of a proper
homothetic motion, it turns into the semidirect sum of 
the loop algebra (\ref{homo5}) and the Witt algebra (\ref{homo7}).
We note that in general the algebra has on $A_\mu$ the above form 
only up to gauge transformations and on-shell trivial symmetries.

{\em D3 and D1+D5 backgrounds.}
We treat now two particularly interesting curved 
backgrounds and
give the symmetry transformations before gauge fixing. 

First we consider a D3-brane supergravity background
with target space metric and 2-form given by
\begin{eqnarray}
& ds^2=H^{-1/2}\eta_{ab} dx^a dx^b + H^{1/2}\delta_{AB} dx^A dx^B & 
\nonumber \\
& b_{mn}=0, \quad H=1 + (R/r)^{4} &
\label{D3back}
\end{eqnarray}
where $r^2=\delta_{AB}x^A x^B$,
$a=0,\dots,3$ and $A=4,\dots,9$.
The rigid symmetries are obtained from (\ref{SOL}) by solving 
(\ref{homo1}) and (\ref{homo2}). Due to $b_{mn}=0$,
the solution
of (\ref{homo2}) is trivial, i.e.\ we can choose $q_{mi}=0$ without
loss of generality. An analysis of (\ref{homo1}) shows that
in this case we have $K_i=0$, i.e.,
there is no proper homothetic motion. Hence, the solutions
of (\ref{homo1}) are exhausted by the Killing vector fields
of the metric in (\ref{D3back}). The latter correspond to
Poincar\'e transformations in the 4-space parallel to the 
D3 brane, and rotations in the transverse directions. The symmetry
transformations of $\varphi$ and $x^m$ read thus in this case
\begin{eqnarray}
\Delta x^a &=& \lambda^a(\varphi) + \lambda^{ab}(\varphi)\eta_{bc}x^c,
\quad \lambda^{ab}=-\lambda^{ba}
\nonumber \\
\Delta x^A &=& \lambda^{AB}(\varphi)\delta_{BC}x^C,\quad
\lambda^{AB}=-\lambda^{BA}
\nonumber\\
\Delta \varphi &=& 0.
\label{sol3}
\end{eqnarray}
The transformations of $A_\mu$ are then obtained from (\ref{SOL}).
(\ref{sol3}) implies that the symmetry group is
in this case a loop version of $ISO(1,3)\times SO(6)$.

Next we discuss the near horizon geometry of (\ref{D3back})
due to its importance for the conjectures in \cite{m}.
Close to the horizon ($r\rightarrow 0$) one can neglect
the constant in the harmonic function $H$ and ends up with 
\begin{equation}
(ds^2)_{hor.} = \frac{r^2}{R^2}\,\eta_{ab}dx^a dx^b +
\frac{R^2}{r^2}\,\delta_{AB}dx^A dx^B\ .
\label{horD3}\end{equation}
Again one finds that the solutions of (\ref{homo1}) 
are exhausted by the Killing vector fields.
However, the asymptotic metric has more isometries than
the original one,
\begin{eqnarray}
\Delta x^a &=& \lambda^a(\varphi)+\lambda^{ab}(\varphi)\eta_{bc}x^c 
+\lambda_D(\varphi)x^a
\nonumber\\
& &-2\lambda_S^b(\varphi)\eta_{bc} x^a x^c 
+ \lambda_S^a(\varphi)(\eta_{bc}x^bx^c+R^4r^{-2})
\nonumber\\
\Delta x^A &=& [2\lambda_S^a(\varphi)\eta_{ab} x^b
-\lambda_D(\varphi)]\, x^A+
\lambda^{AB}(\varphi)\delta_{BC}x^C
\nonumber\\
\Delta \varphi &=& 0.
\label{solD3h}\end{eqnarray}
The additional isometries, corresponding to $\lambda_D$ and $\lambda_S$,
are indeed reminiscent of dilatations and
special conformal transformations in (1+3)-dimensional
 flat space.
The symmetry group is now a loop version of $SO(2,4)\times SO(6)$.
This symmetry enhancement originates from the anti-de Sitter geometry
and corresponds to the supersymmetry enhancement
discussed in~\cite{kallosh}.

Finally, we consider the near horizon
geometry of a D1+D5 supergravity background. The target
space metric and 2-form are given by
\begin{eqnarray}
 ds^2&=&\frac{r^2}{R_1R_5}\, \eta_{\mu\nu}dx^\mu dx^\nu
+\frac{R_1}{R_5}\, \delta_{ab}dx^a dx^b
\nonumber\\
& & +\frac{R_1R_5}{r^2}\,\delta_{AB}dx^A dx^B  
\nonumber\\
b&=&\frac{r^2}{R_1^2} dx^0 \wedge dx^1
+2R_5^2\sin^2\theta_1\sin\theta_2\theta_3
d\theta_1 \wedge d\theta_2
\label{metricD5}\end{eqnarray}
where $r^2=\delta_{AB}x^A x^B$, 
$\mu=0,1$, $a=2,\dots,5$, $A=6,\dots,9$ and the $\theta_i$
are spherical coordinates for the $x^A$ as in \cite{k2}.
Again there are no proper homothetic motions, i.e.\ the
solutions of (\ref{homo1}) are exhausted by the Killing vector fields
of the metric in (\ref{metricD5}). 
The 2-form $b$ is not invariant under all these isometries
but it is still invariant up to exact forms,
as required by (\ref{homo2}). The symmetries form
a loop version of $SO(2,2)\times SO(4)\times ISO(4)$
through
\begin{eqnarray}
\Delta x^\mu &=& \lambda^{\mu}(\varphi)+\lambda^{\mu\nu}(\varphi)
\eta_{\nu\varrho}x^{\varrho} 
+\lambda_D(\varphi)x^{\mu}
\nonumber\\
& & + \lambda_S^{\nu}(\varphi)
[\delta_\nu^\mu(\eta_{\varrho\sigma}x^{\varrho}x^\sigma
+R_1^2R_5^2r^{-2})
-2\eta_{\nu\varrho} x^{\mu} x^{\varrho}]  
\nonumber\\
\Delta x^A &=& [2\lambda_S^{\mu}(\varphi)\eta_{\mu\nu}x^{\nu} -
\lambda_D(\varphi)]x^A + \lambda^{AB}(\varphi)\delta_{BC}x^C
\nonumber\\
\Delta x^a &=& \lambda^a(\varphi) + \lambda^{ab}(\varphi)\delta_{bc}x^c,
\ \Delta \varphi= 0
\end{eqnarray}
where $\lambda^{mn}=-\lambda^{nm}$. The corresponding transformations 
$\Delta A_\mu$ are obtained from (\ref{SOL}), with 
$q_{\mu i}=q_{a i}=0$ and
\begin{eqnarray}
& \lambda^iq_{Ai} =
2\lambda_S^\nu \epsilon_{\nu\mu}x^\mu\delta_{AB}x^B R_5^2r^{-2} &
\nonumber\\
& +\lambda^{BC}x^D
(b_{AB}\delta_{CD}-\frac 12\epsilon_{ABCD}R_5^2r^{-2}).&
\end{eqnarray}

{\em 2d conformal field theories.}
We now discuss the interacting conformal field
theories obtained in the static gauge
$x^\mu=\sigma^\mu$ ($\mu=0,1$)
for world-sheet diffeomorphisms.
Before eliminating the auxiliary fields 
$\gamma_{\mu\nu}$ and $\varphi$,
the action in the static gauge is thus a functional of
\[ \{\phi\}= \{A_\mu,\gamma_{\mu\nu},\varphi,x^2,x^3,\dots\}.\]
This action is of course not invariant
anymore under the transformations $\Delta$ given above.
Rather it is invariant
under particular combinations of these
transformations and compensating world-sheet diffeomorphisms 
preserving the static gauge. These combinations are
\begin{equation}
\delta\phi={\cal L}_\epsilon\phi-[\Delta\phi]_{x^\mu=\sigma^\mu}\ ,\quad
\epsilon^\mu=[\Delta x^\mu]_{x^\mu=\sigma^\mu}
\label{delta}\end{equation}
where ${\cal L}_\epsilon$ is the world-sheet Lie derivative
along $\epsilon^\mu$.
The algebra of the $\delta$'s coincides with the algebra
of $\Delta$'s. Hence, only the realization of these
symmetries changes, but not the corresponding symmetry group.

Let us now illustrate this procedure
for the near horizon D3-brane supergravity background
(\ref{horD3}). The corresponding
action (\ref{density}) reads in the static gauge
\begin{eqnarray}
 S &=&  \frac{1}{2} \int d^2\sigma\left\{
\sqrt{\gamma}\, \gamma^{\mu\nu}
f(\varphi)\, r^2R^{-2}[\eta_{\mu\nu}
+\delta_{\hat{a}\hat{b}}\partial_{\mu}
x^{\hat{a}}\partial_{\nu}x^{\hat{b}} 
\right. \nonumber \\
& & \left. +R^4r^{-4}\delta_{AB}\partial_{\mu}x^A
\partial_{\nu}x^B] + \epsilon^{\mu\nu}D(\varphi)\, F_{\mu\nu}
\right\} 
\label{density1}
\end{eqnarray}
where $\hat{a},\hat{b}=2,3$ correspond to the parallel D3-brane
directions which have not been gauge fixed.
The symmetries of (\ref{density1}) are now obtained
from (\ref{delta}) using (\ref{SOL}) and (\ref{solD3h}). 
For instance, a dilatation symmetry corresponding to
$\lambda_D$ involves a compensating diffeomorphism
with parameter $\epsilon_D^\mu= \lambda_D(\varphi)\sigma^\mu$
and is now realized by
\begin{eqnarray}
\delta_D x^{\hat a}&=&
\epsilon_D^\mu\partial_\mu x^{\hat a}-\lambda_D(\varphi)x^{\hat a}
\nonumber\\
\delta_D x^A&=&
\epsilon_D^\mu\partial_\mu x^A +\lambda_D(\varphi)x^A
\nonumber\\
\delta_D \varphi&=&
\epsilon_D^\mu\partial_\mu\varphi
\nonumber\\
\delta_D \gamma_{\mu\nu}&=&
\epsilon_D^\varrho\partial_\varrho\gamma_{\mu\nu}
+\gamma_{\varrho\nu}\partial_{\mu}\epsilon_D^\varrho
+\gamma_{\mu\varrho}\partial_{\nu}\epsilon_D^\varrho
\nonumber\\
\delta_D A_\mu&=&
\epsilon_D^\varrho\partial_\varrho A_\mu
+ A_\varrho \partial_{\mu}\epsilon_D^\varrho
\nonumber\\
& & -\lambda'_D(\varphi)f(\varphi)[D'(\varphi)]^{-1}
\sqrt \gamma\, \epsilon_{\mu\nu} \gamma^{\nu\varrho} 
R^{-2}r^2\times
\nonumber\\
& &
[\eta_{\varrho\sigma}\sigma^\sigma
+\delta_{\hat a\hat b}x^{\hat a}\partial_\varrho x^{\hat b}
-r^{-4}R^4\delta_{AB}x^A\partial_\varrho x^B].
\label{D}
\end{eqnarray}
These transformations generate symmetries
of (\ref{density1}) for any choice of $\lambda_D(\varphi)$.
This includes dilatations of the standard form for the special
choice $\lambda_D=1$, 
\begin{equation}
\lambda_D=1:\quad
\delta_D \phi=\sigma^\mu\partial_\mu \phi+w(\phi)\phi,
\label{weyl}\end{equation}
where the Weyl weights $w(\phi)$ are given by
\[
w(x^{\hat a})=-1, w(x^A)=w(A_\mu)=1, 
w(\varphi)=0, w(\gamma_{\mu\nu})=2.
\]
Analogously one determines the other 
symmetries in the static gauge. Altogether they form, as before,
a loop generalization of $SO(2,4)\times SO(6)$ with
a loop version of conformal $SO(2,2)$
as a subgroup. This subgroup corresponds to $\lambda^\mu$,
$\lambda^{\mu\nu}$, $\lambda_D$ and $\lambda^\mu_S$, and
the parameters of the compensating 
world-sheet diffeomorphisms for this subgroup are thus
\begin{eqnarray}
\epsilon^\mu_C&=&
\lambda^{\mu}(\varphi) + 
\lambda_D(\varphi)\sigma^{\mu} 
+ [\lambda^{\mu\nu}(\varphi)
- 2\sigma^{\mu}\lambda^{\nu}_S(\varphi)]\eta_{\nu\varrho}\sigma^{\varrho}
\nonumber\\
& &
+\lambda_S^{\mu}(\varphi)(\sigma^{\nu}\sigma^{\varrho}\eta_{\nu\varrho}
+x^{\hat a}x^{\hat b}\delta_{\hat{a}\hat{b}}
+R^4r^{-2}).
\end{eqnarray}
The corresponding conformal transformations of
$x^{\hat a}$, $x^A$ and $\varphi$ can be written compactly as
\begin{eqnarray}
& \delta_C x^{\hat a}=
\epsilon_C^\mu\partial_\mu x^{\hat a}
-\frac{1}{2}(\partial_{\mu}^{exp.}\epsilon^{\mu}_C)x^{\hat a} &
\nonumber\\
& \delta_C x^A=
\epsilon_C^\mu\partial_\mu x^A 
+\frac{1}{2}(\partial_{\mu}^{exp.}\epsilon^{\mu}_C)x^A &
\nonumber\\
& \delta_C \varphi=
\epsilon_C^\mu\partial_\mu\varphi &
\label{C}\end{eqnarray}
where
$\partial_{\mu}^{exp.}$ denotes differentiation only with 
respect to explicit $\sigma^{\mu}$. Note that even the
zero modes of the special
conformal transformations ($\lambda^\mu_S=constant$) 
are nonlinearly realized.

	If we consider (\ref{density1}) in the Born-Infeld action case
and expand in low velocities we get
\begin{eqnarray}
L_{BI} & = &\frac{r^2}{R^2} + \frac{r^2}{2R^2}\,
\delta_{\hat a\hat b}\partial^{\mu}x^{\hat{a}}\partial_{\mu}x^{\hat{b}} 
\nonumber \\ 
 & & 
+\frac{R^2}{2r^2}\,\delta_{AB}\partial^{\mu}x^A\partial_{\mu}x^B 
+\frac{R^2}{4r^2}F^{\mu\nu}F_{\mu\nu} + \ldots 
\label{low}
\end{eqnarray}
where $\mu,\nu$ are raised with $\eta^{\mu\nu}$.

The case of a D-string in the near horizon D1+D5 supergravity background
(\ref{metricD5}) is treated analogously. The resulting
symmetry transformations establish a loop generalization of the
conformal
$SO(2,2)\times SO(4)\times ISO(4)$ symmetry found in \cite{k2}.
The Weyl weights are again easily obtained from the special
dilatation with $\lambda_D=1$ which has again the form (\ref{weyl}) 
and yields
\[
w(x^A)=w(A_\mu)=1, 
w(x^a)=w(\varphi)=0, w(\gamma_{\mu\nu})=2.
\]

{\em Comments.}
The symmetries of D-string actions described above may be
viewed as generalizations of the familiar target space symmetries
of the string. There are two 
important differences to the string case which
are both direct consequences of the presence
of the Born-Infeld gauge field.
First, each target space symmetry gives rise to a family of infinitely 
many symmetries of the D-string action, whereas it yields only one
rigid symmetry of the (Nambu-Goto or Polyakov) string action. 
Second, there is an additional
infinite family of symmetries of the D-string action if the
target space metric admits a proper homothetic motion.
The latter are dilatational symmetries without any counterpart 
in the string case (see \cite{ours1} for an example).

We stress that all these infinitely many symmetries are present
{\em in addition} to the world-sheet symmetries and must not be
confused with the latter. Indeed, the action (\ref{density}) is 
of course also gauge invariant
both under world-sheet diffeomorphisms and under Weyl-transformations 
of $\gamma_{\mu\nu}$, as its string counterpart, 
the Polyakov action. In particular, one may consider the action 
(\ref{density}) in a conformal gauge
for these world-sheet symmetries (rather than in
the static gauge considered above). That action has
infinitely many conformal world-sheet symmetries on top of the
symmetries discussed above. In particular it may thus serve as
a starting point for quantization, along the lines of string 
quantization based on the Polyakov action in a conformal gauge.

{\em Acknowledgements.}
We thank Paul Townsend, Antoine van
Proeyen and J.M. Mart\'{\i}n Senovilla
for discussions.
This work was supported in part by NSF grant PHY-9511632, the Robert A.
Welch Foundation AEN95-0590 (CICYT), GRQ93-1047 (CIRIT) and
by the Commission of European Communities CHRX93-0362(04).
FB was supported by the Deut\-sche For\-schungs\-ge\-mein\-schaft.

\end{document}